\documentclass[twoside,twocolumn,showpacs,floatfix]{revtex4}
\usepackage[T1]{fontenc}
\usepackage{textcomp}
\usepackage{times}
\usepackage{dcolumn}
\usepackage{amsmath,amsfonts,amssymb}
\usepackage[dvips]{graphicx,color}
\newcommand{\REM}[1]{}
\begin{document}
\title{First order phase transitions in classical lattice gas spin models}
\author{H. Chamati}
\affiliation{Institute of Solid State Physics,
Bulgarian Academy of Sciences, \\
72 Tzarigradsko Chauss\'ee, 1784 Sofia, Bulgaria.}
\author{S. Romano}
\affiliation{Unit\'a di Ricerca CNISM e
Dipartimento di Fisica "A. Volta", Universit\'{a} di Pavia
via A. Bassi 6, I-27100 Pavia, ITALY
}
\date{}

\begin{abstract}
The present paper considers some classical
ferromagnetic lattice--gas models,
consisting of particles that carry
$n$--component spins ($n=2,3$)
and associated with a $D$--dimensional lattice 
($D=2,3$);
each site can host one particle at most, thus implicitly allowing
for hard--core repulsion; the pair interaction, restricted to nearest
neighbors, is ferromagnetic, and site occupation is also
controlled by the chemical potential $\mu$.
The models had previously been investigated 
by Mean Field and Two--Site Cluster treatments (when $D=3$), 
as well as  Grand--Canonical Monte Carlo simulation in the case $\mu=0$,
for both $D=2$ and $D=3$; the obtained results showed 
the same kind of critical behaviour as the one known for their 
saturated lattice counterparts, corresponding to one particle per site.
Here we addressed 
by Grand--Canonical Monte Carlo simulation the case where
the chemical potential is negative and  sufficiently large in magnitude; 
the value $\mu=-D/2$ was chosen 
for each of the four previously investigated counterparts, 
together with  $\mu=-3D/4$ 
in an additional instance.
We mostly found evidence of first order transitions, 
both for $D=2$ and $D=3$,
and quantitatively characterized their behaviour.
Comparisons  are also made with recent
experimental results.


\end{abstract}

\maketitle
\section{Introduction}

There exist a few statistical mechanical models
involving classical continuous spins, by now  extensively 
investigated, and 
{which  play} a central r\^ole in a 
variety of real physical 
situations; they have been especially studied in their saturated lattice (SL)
version, where each lattice site hosts one spin. 
The interest in their lattice gas (LG) extensions
was recently  revived,  and they were addressed by means of analytical
theories and simulation (see for example
\cite{romano1999,sokolovskii2000,romano2000,maciolek2004,chamati2005a,chamati2005b,raddlg,chamati2006}
and references therein).
LG spin models are obtained from 
{their SL
counterparts by allowing for fluctuations of occupation numbers, 
also controlled by the chemical potential $\mu$.} 
These models have often been used in
connection with alloys and absorption; the methodology somehow
allows for pressure and density effects.

As for symbols and
definitions, classical SL spin models involve $n-$component unit
vectors ${\bf u}_k$, associated with a $D-$dimensional (bipartite)
lattice $\mathbb{Z}^D$; let ${\bf x}_k$ denote dimensionless
coordinates of the lattice sites, and let $u_{k,\alpha}$ denote
cartesian spin components with respect to an orthonormal basis ${\bf
e}_{\alpha}$, whose unit vectors can be taken as defined by the
lattice axes. 
The orientations of the magnetic
moments of the particles
are parameterized by usual polar
angles $\{\phi_j\}$ ($n=2$) or spherical ones
$\{(\varphi_j,~\theta_j)\}$ ($n=3$). The interaction potential,
restricted to nearest neighbors, is assumed to be ferromagnetic and,
in general, anisotropic in spin space, i.e.
\begin{eqnarray}\label{eq01-a}
\Phi_{jk}=\epsilon Q_{jk},~Q_{jk}= - 
\left[
a u_{j,n} u_{k,n} + b \sum_{\alpha < n} u_{j,\alpha} u_{k,\alpha}
\right];~ \nonumber\\
\epsilon >0,~a \geq 0,~b \geq 0,~a+b>0,~\max(a,b)=1.
\end{eqnarray}
{Notice also that the condition $\max(a,b)=1$
in the above equation can always be satisfied by a suitable 
rescaling of $\epsilon$; 
here and in the following the quantity $\epsilon$ will be used 
to set temperature and energy scales;
thus $T = k_B t /\epsilon$, where $t$ denotes the
absolute temperature and $k_B$ 
is the Boltzmann constant; the corresponding 
(scaled) Hamiltonian is given by:}
\begin{equation}
\Lambda=  \sum_{\{j<k\}}Q_{jk}.
\end{equation}

The case $n=1$ corresponds to the Ising model; 
isotropic $O(n)$-symmetric models ($n >1$) correspond to $a=b$, 
$Q_{jk}=- {\bf u}_j\cdot {\bf u}_k$,
and are referred to as planar rotators (PR, $n=2$) or classical 
Heisenberg model (He, $n=3$); the extremely anisotropic and 
$O(2)$-symmetric XY model is defined by $n=3$, $a=0$.
For these models
the simplification resulting from the neglect of translational
degrees of freedom makes it possible to obtain
rigorous mathematical results \cite{sinai1982,georgii1988,rBruno}
entailing existence or absence of a
phase transition, and, on the other hand,
to study it by a whole range of techniques,
such as Mean Field (MF) and Cluster Mean Field treatments,
high-temperature series expansion of the partition
function, Renormalization Group
(for a recent review see \cite{pelissetto2002}),
computer simulation (usually via Monte Carlo (MC) methods
\cite{newman1999}).

LG extensions of the continuous--spin
potential model considered here are defined by Hamiltonians
\begin{equation}\label{eq02}
\Lambda = \sum_{\{j<k\}} \nu_j \nu_k (\lambda - \Omega_{jk})- \mu N
,~\qquad N=\sum_k \nu_k,
\end{equation}
where $\nu_k=0,1$ denotes occupation numbers; notice that $\lambda
\leq 0$ reinforces the orientation--dependent term, whereas $\lambda
>0$ opposes it, and that a finite value of $\lambda$ only becomes
immaterial in the SL limit $\mu \rightarrow + \infty$.
It is worth mentioning that in such systems the fluctuating occupation numbers
give rise to additional fluid-like observables in comparison to
the usual SL situation.

Rigorous results entailing existence or absence 
of an ordering transition are also known for LG models 
with continuous spins \cite{rAZ04,rAZ01,rAZ02,rAZ03,rCSZ}. 
For some models defined by $D=3$, interactions 
isotropic in spin space, 
and supporting a ferromagnetic phase transition in their SL version, 
it has been proven that there exists a 
$\mu_0$, such that, for all $\mu > \mu_0$, the system supports a
ferromagnetic transition,
with a $\mu-$dependent transition temperature.
Notice that $\mu_0<0$
when $\lambda \le 0$ \cite{rAZ01,rAZ02,rAZ03}, whereas a positive 
$\mu_0$ may be needed when $\lambda >0$.
More recently \cite{rCSZ}, the existence of a first-order transition,
involving discontinuities in both density and magnetization,
has been proven for the isotropic case (and $D=3$), in a
suitable r\'egime of low temperature and negative $\mu$. 

For $D=2$, the SL--PR  model produces at low--temperature the 
extensively studied 
Berezinski\v\i-Kosterlitz-Thouless (BKT) transition \cite{rKT0,rBKTrev3};
the existence of such a transition for the LG counterpart
has been proven rigorously as well \cite{rGTZ}.
More recently, it was rigorously proven \cite{rERZ} that, for
$\mu$ negative and sufficiently large in magnitude,
the transition becomes first--order.

Notice also that the above mathematical theorems 
do not yield useful numerical estimates of the $\mu$ value 
where the change of transition sets in; some answer to this
question can be looked for by analytical approximations
such as MF or Two--Side--Cluster (TSC) treatments 
\cite{romano2000,chamati2005a}, or by simulation 
\cite{romano2000,chamati2005a,chamati2006}. 

The Hamiltonian (Eq. (\ref{eq02})) can be interpreted as describing
a two--component system consisting of interconverting ``real''
($\nu_k=1$) and ``ghost'', ``virtual'' or ideal--gas particles
($\nu_k=0$); both kinds of particles have the same kinetic energy,
$\mu$ denotes the excess chemical potential of ``real'' particles
over ``ideal'' ones, and the total number of particles equals the
number of available lattice sites (semi--Grand--Canonical
interpretation).
The semi--Grand--Canonical interpretation was also used in early
studies of the phase diagram of the two--dimensional planar rotator, 
carried out by the
Migdal--Kadanoff RG techniques, and aiming at two--dimensional
mixtures of $^{3}$He and $^{4}$He \cite{rhe01,rhe03}, where
non--magnetic impurities correspond to $^{3}$He.

In the three--dimensional case, the topology of the phase diagram of
the model (\ref{eq02}) had been investigated by 
MF  and TSC approximations for 
{the  Ising
\cite{sokolovskii2000} as well as PR cases \cite{romano2000} in the
presence of a magnetic field,} and for He at zero magnetic field
\cite{chamati2005a}. These investigations
were later extended  \cite{chamati2005b} 
to extremely anisotropic (Ising--like) two--dimensional LG models 
defined by $a=1,~b=0$ in Eq. (\ref{eq01-a}), and 
in the absence of a magnetic field  
{as well.}
The studied models were found to exhibit a
tricritical behaviour i.e. the ordering transition 
{turned out to be  of first order
for $\mu$ below an appropriate 
threshold, and of  second order above it.} When the transition is of 
first order, the orientationally ordered phase is also denser than the
disordered one. For the three--dimensional PR these finding were
confirmed, recently, by simulation in connection with the phase
diagram of He \cite{maciolek2004}.
{It has been found that, 
despite the simplicity of LG spin models, 
their predictions  broadly agree with the ones obtained by means
of more elaborate magnetic fluid models (see e.g. \cite{omelyan2004}
and references therein).}

{On the other hand, 
thermodynamic and structural properties had  been
investigated by means of  Grand--Canonical Monte Carlo simulations as well
\cite{romano2000,chamati2005a},
for particular 
values of the chemical potential equal or close to zero.}  
It had  been found that there is a {\em second order} 
ferromagnetic phase transition 
manifested by a significant growth of magnetic and density
fluctuations. The transition temperatures were found to be about 20\%
lower than that of the corresponding SL values and the critical behaviour 
of the investigated models to be consistent  with that of their SL
counterparts. Furthermore it had been found that MF yields a
qualitatively correct picture, and the quantitative agreement with
simulation could be improved by TSC, which has the advantage of
predicting two-site correlations.

Notice also that the above Hamiltonian (Eq. (\ref{eq02})) describes
a situation of {\em annealed} dilution; on the other hand,
two--dimensional models in
the presence of {\em quenched} dilution, and hence the effect of
disorder on the BKT transition, have been investigated using the PR
model
\cite{rque01,rque02,leonel2003,berche2003,surungan2005,wysin2005}
and very recently its XY counterpart
\cite{wysin2005}; it was found that a
sufficiently weak disorder does not destroy the transition, which
survives up to a concentration of vacancies close to the percolation
threshold.
Let us also remark that
two--component spins are involved in the PR case, whereas XY
involves three--component spins but only two of their components are
involved in the interaction: in this sense the two models entail
different anchorings with respect to the horizontal plane in spin
space.
Two--dimensional annealed lattice models were investigated
\cite{chamati2006} as well, and the obtained results for $\mu=0$ or 
a moderately negative $\mu$ were
found to support those obtained for quenched models. For a large
negative $\mu$, renormalization group treatments had 
suggested \cite{rhe01,rhe03} that the transition between
the BKT and the paramagnetic phase is of  first order.

In this paper, we present an extensive Monte Carlo study of some
LG ferromagnetic models,
{where $\mu$ is negative and
comparatively large in magnitude}
(notice that $\mu < -D$ would produce an empty ground--state), 
in order to gain insights into their critical behaviour and to check
the impact of the chemical potential on their physical properties. On
the other hand, for $D=3$, 
we will also  test the MF or TSC approximations used
to obtain the phase diagrams of Refs. \cite{romano2000,chamati2005a}. 
In keeping with our previous studies, the models are further simplified by
choosing $\lambda=0$, i.e. no pure positional interactions.
As for the values of the chemical potential, we chose
$\mu=-D/2$ in the four cases corresponding to our
previous investigations with $\mu=0$, and carried out additional simulations
for $D=2$, PR and $\mu=-3D/4$, as explained below.

{The rest of the paper is organized as follows: 
the  simulation procedure is 
briefly explained 
in Section \ref{simulation}, 
section \ref{results} analyzes  the 
simulation results. Finally, the effects caused by the
chemical potential on the nature of the transition are  discussed
in Section \ref{conclusion}, which summarizes our
results, and where some comparisons are made with a recent
experimental work.}

\section{Monte Carlo simulations}\label{simulation} A detailed
treatment of Grand--Canonical simulations can be found in or via Refs.
\cite{romano1999,chamati2005a,rsim3}; the method outlined here has
already been used in our previous studies of other LG models
\cite{romano2000,chamati2005a,chamati2005b}. Simulations were carried 
out on periodically repeated samples,
consisting of $V=L^D$ sites,
where  $L=40,80,120,160$ for $D=2$, and $L=10,20,30$ for 
$D=3$, i.e. in keeping with 
the named previous studies of ours;
 calculations were
carried out in cascade, in order of increasing reduced temperature $T$.

The two basic MC steps used here were Canonical and
semi--Grand--Canonical attempts; in addition two other features were
implemented \cite{rmult0,rHR}: (i) when a lattice site was visited,
Canonical or semi--Grand--Canonical steps were randomly chosen with
probabilities ${\cal P}_{\rm can}$ and ${\cal P}_{\rm GC}$,
respectively; we used ${\cal P}_{\rm can}/{\cal P}_{\rm GC}=n-1$,
since spin orientation is defined by $(n-1)$ angles, versus one
occupation number and (ii) sublattice sweeps (checkerboard
decomposition) \cite{rmult0,rHR}; thus each sweep (or cycle)
consisted of $2V$ attempts, first $V$ attempts where the lattice
sites was chosen randomly, then $V/2$ attempts on lattice sites of
odd parity, and finally $V/2$ attempts on lattice sites of even
parity. Equilibration runs took between $25 \,000$ and $200\,000$
cycles, and production runs took between
$250\,000$ and $1\,000\,000$;
macrostep averages for evaluating statistical errors were taken over
$1\,000$ cycles. Different random--number generators were used, as
discussed in Ref. \cite{rHR}.

Computed thermodynamic observables included mean Hamiltonian
per site, $H=(1/V)\left< \Lambda \right>$, density
$\rho = (1/V) \left< N  \right>$, as well as their derivatives
with respect to temperature or chemical potential,
$C_{\mu V}/k_B = (1/ V)
(\partial \left< \Lambda \right>/\partial T)_{\mu,V}$,
$\rho_T= (\partial \rho/\partial T)_{\mu,V}$,
$\rho_{\mu}=(\partial \rho /\partial \mu)_{T,V}$,
defined by appropriate fluctuation formulae 
\cite{rsim3}.

We also calculated mean magnetic moment per site and
susceptibility, defined by
\begin{equation}
M =\frac1V \left< \sqrt{{\bf F} \cdot {\bf F}} \right>,
\end{equation}
where for PR or He the vector ${\bf F}$ is defined by 
\begin{equation}\label{eqfl07}
\mathbf{F}=\sum_{k=1}^V \nu_k\mathbf{\bf u}_k,
\end{equation}
whereas only the in--plane components of the vector
spins (i.e. only the Cartesian components explicitly involved
in the interaction potential) are accounted for in the XY case.

The behaviour of the susceptibility was  investigated by considering the
two quantities:
\begin{equation}\label{eqchi1}
\chi_1=\frac{\beta}{V}
\left(\left<\mathbf{F}\cdot\mathbf{F}\right>
- \left\langle|\mathbf{F}| \right\rangle^2 \right)
\end{equation}
and
\begin{equation}\label{eqchi2}
\chi_2 = \frac{\beta}{V}
\left\langle\mathbf{F}\cdot\mathbf{F}\right\rangle;
\end{equation}
simulation estimates of the susceptibility 
\cite{newman1999,paauw1975,peczak1991} 
are defined by
\begin{eqnarray}
\chi = \left\{ \begin{array}{ll}
\chi_1,& \ \ \ \mathrm{in~the~ordered~region} \\
\chi_2,& \ \ \ \mathrm{in~the~disordered~region}
\end{array}
\right. ;
\label{eqchi4}
\end{eqnarray}
notice also that, for a finite sample, $\chi_2 \leq \beta V$,
and that $\chi=\chi_2$ in two dimensional cases.

A sample of $V$ sites contains $D V$ distinct
nearest--neighbouring pairs of lattice sites; we worked out pair
occupation probabilities, i.e. the mean fractions $R_{JK}$ of pairs
being both empty ($R_{ee}=\left<(1-\nu_j)(1-\nu_k)\right>$), both
occupied ($R_{oo}=\left<\nu_j\nu_k\right>$), or consisting of an
empty and an occupied site
($R_{eo}=\left<(1-\nu_j)\nu_k+(1-\nu_k)\nu_j\right>$). It should be
noted that $R_{ee}+R_{oo}+R_{eo}=1$.

Short-- and long--range positional correlations were compared by
means of the excess quantities
\begin{equation}
R^{\prime}_{oo}=\ln \left( \frac{R_{oo}}{\rho^2} \right),~
R^{\prime \prime}_{oo}=R_{oo}-\rho^2,
\label{eqexcpos}
\end{equation}
collectively denoted by $R^*_{oo}$ (notice that these two
definitions entail comparable numerical values).

Quantities such as $\rho$, $\rho_T$, $\rho_{\mu}$ and the above pair
correlations $R_{JK}$ or $R^*_{oo}$ can be defined as ``fluid--like'',
in the sense that they all go over the trivial constants in the SL
limit. Let us also remark \cite{chamati2006}
that some of the above definitions (e.g.
$C_{\mu,V}$ and $\rho_T$) involve the total potential energy both in
the stochastic variable and in the probability measure (``explicit''
dependence), whereas some other definitions, e.g. $\rho_{\mu}$ or
the quantities $R_{JK}$, involve the total potential energy only in
the probability measure (``implicit'' dependence).

\section{Simulation results}\label{results}

\subsection{$D=2$, PR, $\mu=-1$}
\begin{figure}[h!]
\centerline{
\resizebox{\columnwidth}{!}{\includegraphics{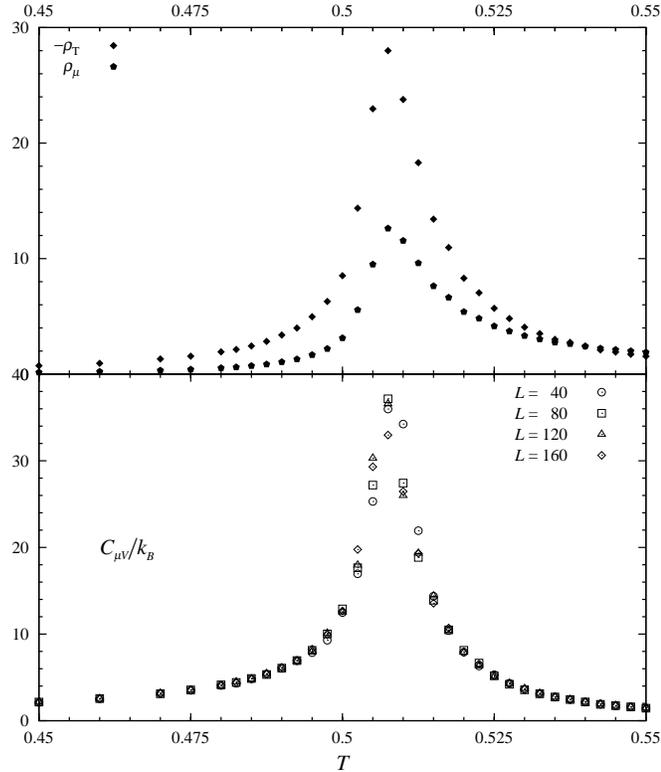}}}
\caption{Simulation estimates for
the specific heat per site $C_{\mu V}$ versus temperature,
obtained with different sample sizes, for the two--dimensional LG--PR and 
$\mu=-1$. Simulation
results for $\rho_T$ and $\rho_{\mu}$ obtained
with the largest examined sample size are shown on the top.
Statistical errors range between 1 \% and 5 \%.}
\label{d2n2mu100der}
\end{figure}

Simulation results, obtained in the named cases
for a number of
observables, such as the mean energy per site and density, showed that
these quantities evolve with the temperature in a smooth way, 
and were
found to be independent of sample sizes. 
In comparison with Ref. \cite{chamati2006},
{their temperature derivatives 
$C_{\mu,V}$ and $\rho_T$ (Fig. \ref{d2n2mu100der}) 
showed 
recognizably more
pronounced peaks} about the same temperature 
$T\approx0.51$, around which the sample size
dependence of results became slightly more pronounced.
Comparison  with our previous results shows that
the location of the maximum of $C_{\mu V}$ is shifted towards lower 
temperatures as $\mu < 0$  grows in magnitude.

\begin{figure}[h!]
\centerline{
\resizebox{\columnwidth}{!}{\includegraphics{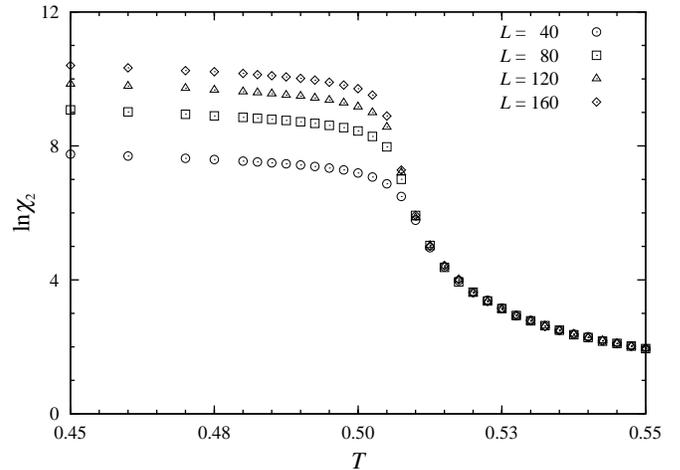}}}
\caption{Simulation estimates for
the logarithm of the magnetic susceptibility $\chi_2$ versus temperature,
obtained with different sample sizes, for the two--dimensional LG--PR 
and $\mu=-1$. 
Unless otherwise stated, here  and in the following figures,
 statistical errors fall within symbol sizes.}
\label{d2n2mu100susc}
\end{figure}

Plots of $\ln \chi_2$ versus $T$, reported in Fig. \ref{d2n2mu100susc},
show results independent of sample size
for $T \gtrsim 0.52$, and then their  pronounced increase with sample 
size for $T \lesssim 0.51$, suggesting its divergence with $L$.
In general this case \textit{qualitatively} reproduces our previous 
simulation results
\cite{chamati2006}, but with more pronounced derivatives and peaks at
a lower temperature. In order to estimate the critical temperature we
applied the finite size scaling analysis, along the lines discussed
in reference \cite{chamati2006},
and  here again we found a BKT transition occurring at
$T_{BKT}=0.502\pm0.002$, corresponding to a particle density about
$0.832\pm0.003$.
{Comparison between  our previous results
and the present ones shows 
that both transition temperature and  ``critical'' particle concentration are
monotonically increasing with the chemical potential  
(see Table \ref{table01}).}

\begin{table*}
\caption{Transition temperatures $\Theta$ and ``critical''
particle density $\rho_c$ of PR
and XY models for some selected values of the chemical potential $\mu$.
 Depending on $\mu$ we
have either a BKT or a first
order one (I); 
here  $\rho_c$ denotes the density at the 
BKT transition temperature.}
\label{table01}
\begin{ruledtabular}
\begin{tabular}{lrlll}
Model &$\mu$\ \ & Transition  &   $\Theta$       &        $\rho_c$          \\
\hline
PR(n=2)&$\infty$& BKT        &$0.907\pm0.004$\cite{wysin2005}  & 1.             \\
      &$ 0.1$   & BKT        &$0.75\pm0.01$\cite{chamati2006}  &$0.938\pm0.002$ \\
      &$ 0.0$   & BKT        &$0.733\pm0.003$\cite{chamati2006}&
                                       $0.924\pm0.003$\cite{chamati2006}        \\
      &$-0.2$   & BKT        &$0.71\pm0.01$\cite{chamati2006}  &$0.900\pm0.002$ \\
      &$-1.0$   & BKT        &$0.502\pm0.002$     &$0.832\pm0.003$              \\
      &$-1.5$   & I          &$0.279\pm0.001$     &  $-$                        \\
XY(n=3)&$\infty$& BKT        &$0.700\pm0.005$\cite{wysin2005}  & 1.             \\
       &$0.0$   & BKT        &$0.574\pm0.003$\cite{chamati2006}&
                                       $0.918\pm0.004$\cite{chamati2006}        \\
       &$-1.0$  & I          &$0.333\pm0.001$     &  $-$                        \\
\end{tabular}
\end{ruledtabular}
\end{table*}

For the SL--PR model the maximum of the specific heat 
is located at about $15\%$ \cite{tobochnik1979} above the BKT transition;
for the LG--PR model and $\mu=0$ 
\cite{chamati2006} we had found a broad peak about $5\%$ above
the BKT transition, and here  
we find a sharper one about $2\%$ above the transition temperature.

For $\mu=-1$, fluidlike quantities show qualitatively similar
behaviours as their counterparts obtained for $\mu=0$. 
Results for $\rho_T$ and
$\rho_\mu$, obtained with the largest sample sizes are
reported in Fig. \ref{d2n2mu100der}; they
were found to behave in a similar fashion to the specific heat
and to exhibit  sharper peaks taking place at the same temperature as
that of $C_{\mu V}$. Recall that $\rho_\mu$ has a broad maximum for
$\mu=0$. In other words, here the ferromagnetic
orientational fluctuations taking place in the transition range 
do produce stronger fluctuations of site occupation variables,
and this tends to reduce  the difference  between ``implicit''
and ``explicit'' dependencies on the potential energy
as mentioned in Ref. \cite{chamati2006}.

\begin{figure}[h!]
\centerline{
\resizebox{\columnwidth}{!}{\includegraphics{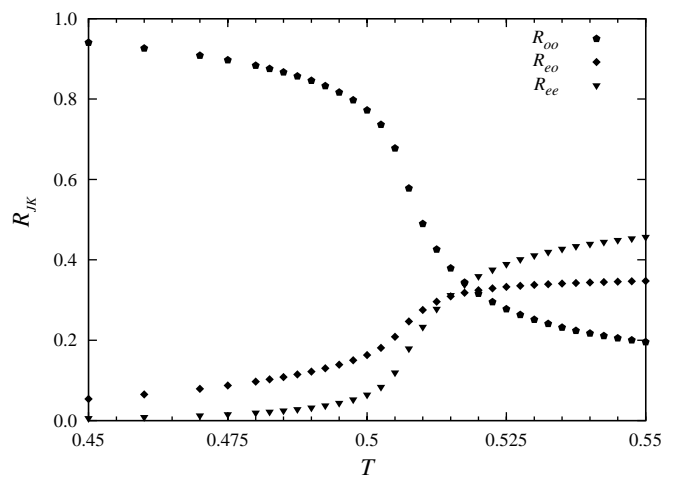}}}
\caption{Simulation estimates for
the pair occupation probabilities $R_{JK}$ 
versus temperature, for the two--dimensional
LG--PR with linear sample size $L=160$. 
The results refer to $\mu=-1$.
}
\label{d2n2mu100pair}
\end{figure}

Pair occupation probabilities $R_{JK}$ were found  to be
insensitive to sample sizes; results for our largest sample
size are shown in Fig. \ref{d2n2mu100pair}. These quantities are
monotonic functions of temperature as their counterparts for $\mu=0$, 
but with more rapid variations
across the transition region,
in accordance with the sharper
maximum of $\rho_\mu$. Their behaviours suggest
inflection points roughly corresponding to the maximum of $\rho_\mu$.

\begin{figure}[h!]
\centerline{
\resizebox{\columnwidth}{!}{\includegraphics{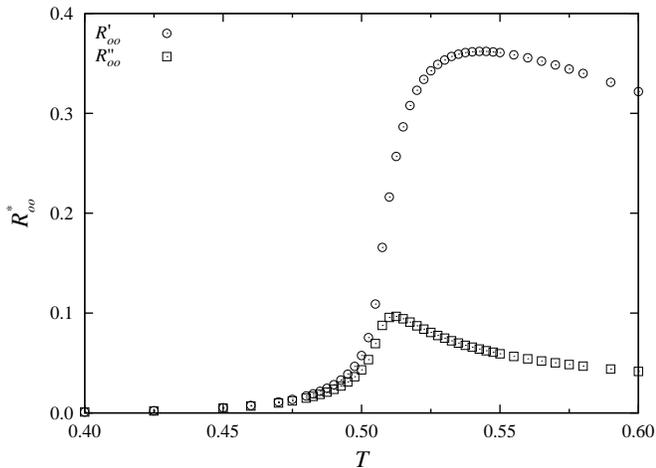}}}
\caption{Simulation estimates for
the excess quantities $R_{oo}^*$ for the two--dimensional LG--PR.
Simulation results were obtained with $L=160$ and $\mu=-1$.
}
\label{d2n2mu100rstar}
\end{figure}

Short-- and long--range positional correlations have been compared
via the excess quantities $R_{oo}^*$, whose simulation results for the
largest sample size are shown in Fig. \ref{d2n2mu100rstar}, showing 
sharper maxima than their counterparts corresponding to $\mu=0$.
Notice also that the position of the maximum
for $R^{\prime \prime}_{oo}$ again corresponds
to the location of the peak of $C_{\mu V}$.
The quantities $R_{oo}^*$ are rather small, and this could be traced back to
the absence of pure positional interactions.

\subsection{$D=2$ and first--order transitions}
\begin{figure}[h!]
\centerline{
\resizebox{\columnwidth}{!}{\includegraphics{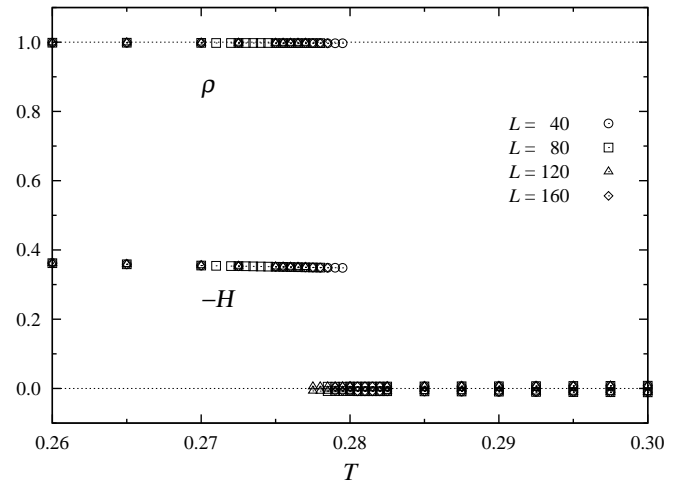}}}
\caption{Simulation results for
the density $\rho$ and the mean energy per site $-H$
obtained for the two--dimensional LG--PR. The value $\mu=-1.5$ was
used for the present simulations.
}
\label{d2n2hyst}
\end{figure}

\begin{figure}[h!]
\centerline{
\resizebox{\columnwidth}{!}{\includegraphics{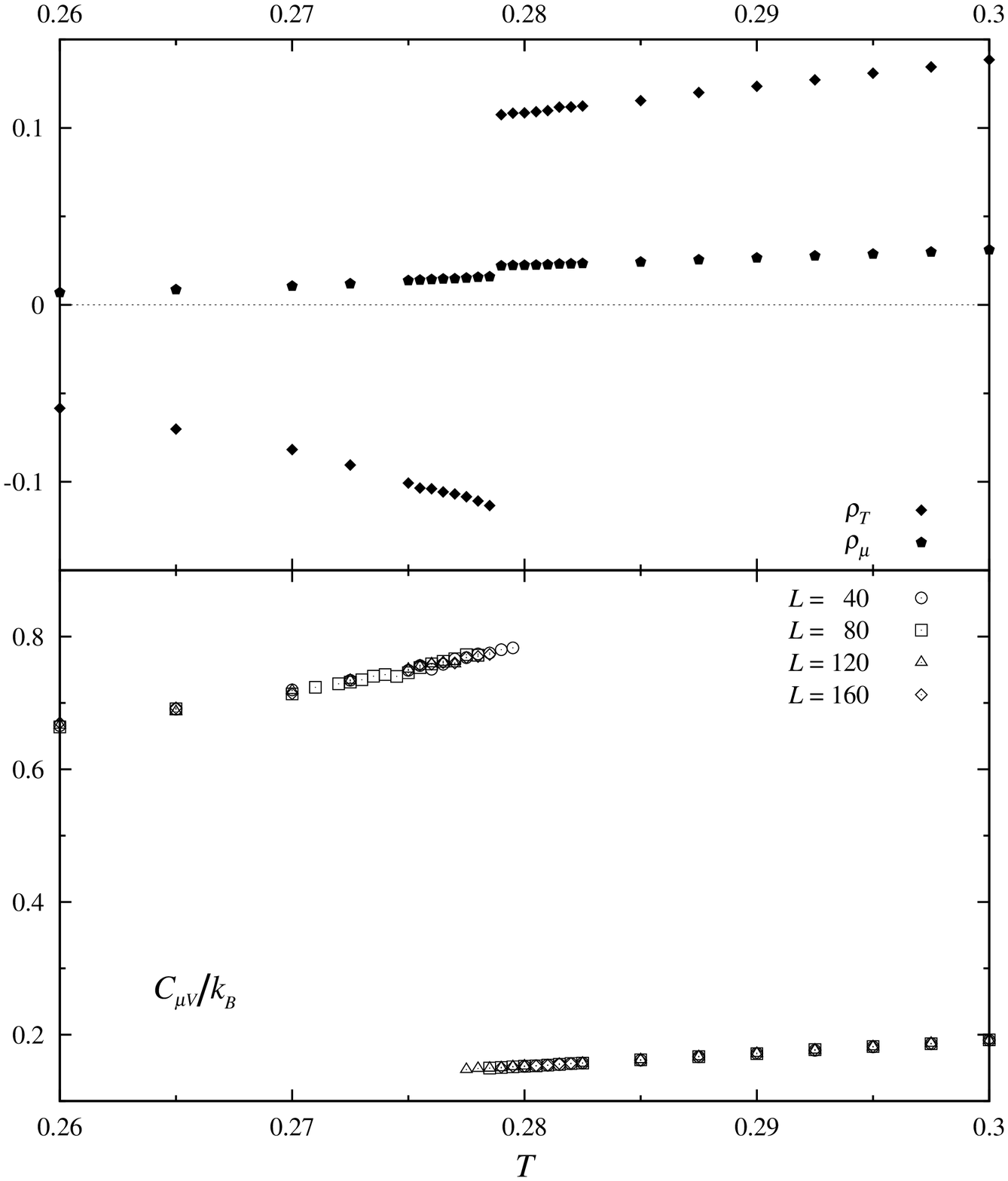}}}
\caption{Simulation estimates for
the specific heat per site $C_{\mu V}$ versus temperature,
obtained with different sample sizes for 
the two--dimensional LG--PR and $\mu=-1.5$. Simulation
results for $\rho_T$ and $\rho_{\mu}$ obtained
with the largest examined  sample size are shown on the top.
Statistical errors range between 1 \% and 5 \%.}
\label{d2n2mu150der}
\end{figure}

\begin{figure}[h!]
\centerline{
\resizebox{\columnwidth}{!}{\includegraphics{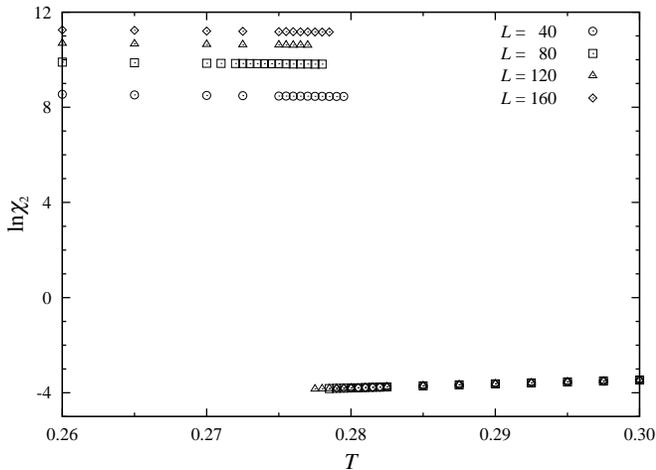}}}
\caption{Simulation estimates for
the logarithm of the magnetic susceptibility $\chi_2$ against temperature,
obtained with different sample sizes for 
the two--dimensional LG--PR and $\mu=-1.5$.}
\label{d2n2mu150susc}
\end{figure}

Additional simulations carried out for $D=2$, PR, $\mu=-3D/4$, 
showed a recognizably different  
{scenario.}
Here, for all investigated sample sizes,
we found pronounced jumps of various
observables, such as $H$, $\rho$ (FIG. \ref{d2n2hyst}) and even $M$
(which kept  decreasing with increasing sample size), taking place over
a narrow temperature range, $\Delta T =0.0005$.
Notice that $\chi_2$ remains independent of sample sizes in the
high--temperature r\'egime, and then develops a pronounced
increase with sample size. From a comparison of the behaviours of
$\chi_2$ for $\mu=-1$ (Fig. \ref{d2n2mu100susc}) and $\mu=-1.5$
(Fig. \ref{d2n2mu150susc})
one can observe the change of the critical behaviour at the two
values of $\mu$. 
For $\mu=-1.5$ the thermodynamic observables show
a discontinuous behaviour characteristic of a first--order 
transition, now to a low--temperature BKT phase.
The behaviours of $C_{\mu V}, \rho_T, \rho_{\mu}$
are shown in Fig. \ref{d2n2mu150der}, and also exhibit
pronounced differences from their counterparts
in the previous case (see also below). 

This result confirms previous RG
predictions \cite{rhe01,rhe03}; 
on the other hand, recent simulation studies 
addressing quenched dilution have found  that the
transition temperature vanishes  
below the percolation threshold 
\cite{leonel2003,berche2003,surungan2005,wysin2005}.

Notice that usage of the Grand--Canonical
ensemble allows quite wide changes of density with temperature;
in the investigated cases we used $\mu > -D$, and found
that $\rho \approx 1$ in the low--$T$ phase, where $\rho_T <0$; 
such  changes are obviously excluded from the start in the treatment of a 
quenched--dilution model.
On the other hand, 
values $\mu<-D$ produce
an 
{essentially} empty ground--state; in this 
r\'{e}gime one can expect that 
$\rho$ to increase with $T$, only becoming appreciable above some threshold, 
and that the BKT phase disappears.

Here and in the following subsection,
transitional properties such as $\Delta H$, $\Delta\rho$, (as well as
$M$, in the next subsection), 
were estimated by analyzing simulation results for
the largest sample size as discussed in Refs.
\cite{rtrans01,rtrans02}. The relevant results are reported
in Table \ref{table02}.

\begin{table}[h!]
\caption[]{
A summary of simulation estimates for properties at first--order
transition for the two--dimensional models obtained using simulations.}
\label{table02}
\begin{ruledtabular}
\begin{tabular}{lllll}
Model & $\mu$  &  $\Theta$       & $\Delta H$ & $\Delta \rho$ 
\\
\hline
PR    & $-1.5$ & $0.279\pm0.001$ & $0.3562\pm0.0005$ & $0.9917\pm0.0001$ 
\\    
XY    & $-1.0$ & $0.332\pm0.001$ &  $0.664\pm0.002$  & $0.910\pm0.001$  
\\
\end{tabular}
\end{ruledtabular}
\end{table}
Let us now turn to the discussion of the nature of the
low--temperature phase. Here the magnetization was found to exhibit a
power--law decay with increasing sample size. A fit to the expression
\begin{equation}
\ln M = -b_1\ln L + b_0, \qquad b_1>0;
\end{equation}
showed that the ratio $b_1(T)/T$ is a constant. This shows that the
magnetization goes to zero in the thermodynamic limit 
($L\to\infty$),  as predicted by the Mermin Wagner theorem for $2D$
systems, where no long range order should survive. Note that this
behaviour is consistent with the spin wave theory developed for the
two--dimensional saturated planar rotator model
\cite{tobochnik1979,archambault1997}. 

Results for $\ln\chi_2$ against
temperature (Fig. \ref{d2n2mu150susc}) were found to be independent of
sample size when $T \gtrsim 0.281$, and showed a recognizable increase 
with it 
{(a linear dependence of 
$\ln\chi_2$ on $\ln L$) when 
$T \lesssim 0.278$.
Thus in the low--temperature
region the susceptibility 
exhibits  a power law divergence with  the linear sample size,  
showing a BKT phase \cite{tobochnik1979,archambault1997}.} 

As for simulation results obtained for the XY LG model with $\mu=-1$,
it was found that the thermodynamic
quantities have qualitatively similar behaviours as those obtained for
the above LG--PR  with $\mu=-1.5$. The phase transition was found to be 
first order taking
place at $T=0.332\pm0.001$; estimates
of transition temperatures reported in Table \ref{table01}
show that they increase as 
a function of the chemical potential.
Transitional properties of the mean energy, the
density and the magnetization are presented in Table \ref{table02}.
Here again we have found that the transition 
{takes place from a paramagnetic to a BKT--like phase.}

Let us emphasize that, as remarked above, PR and XY models entail
different anchorings with respect to the horizontal plane in spin
space; this difference correlates with the pronounced qualitative
different in transition behaviour observed  
when $\mu=-1$. 

{When  both PR and XY lattice gas models exhibited 
a first order phase 
transition,  
their fluidlike quantities 
were found to behave in a
qualitatively similar way. The following  
 discussion  will  concentrate on these properties for 
the XY model.} 
\REM{The quantities
$\rho_T$ and $\rho_\mu$ are increasing against the temperature in the
low--temperature region and remains constants at temperatures higher
than the transition temperature.} 

Fig. \ref{d2n2mu150der} shows that $\rho_T$
is negative  and decreases
with increasing $T$ in the low--temperature region
(where it is essentially driven by orientational correlations),
and then it becomes weakly positive and increasing with $T$
in the high--temperature phase; thus, here and in the following subsection, 
$\rho$ decreases with $T$ in the low--temperature phase, and then
increases with $T$ in the high--temperature region.
On the other hand, here  $\rho_{\mu}$ 
is  an increasing function of $T$, exhibiting a 
jump across the transition.

\begin{figure}[h!]
\centerline{
\resizebox{\columnwidth}{!}{\includegraphics{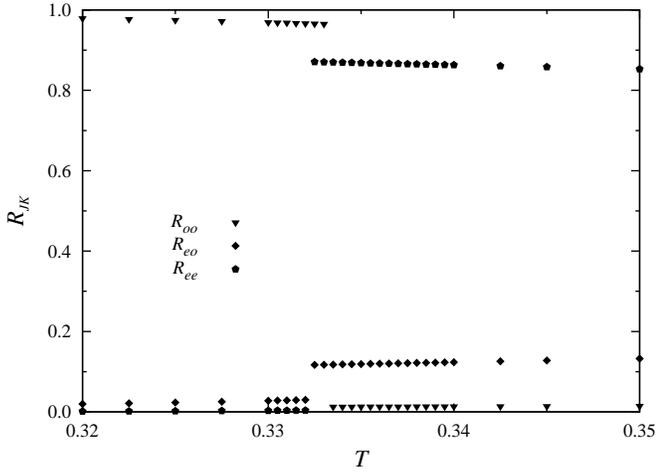}}}
\caption{
Simulation estimates for the three pair occupation
probabilities $R_{JK}$.
for the two--dimensional
LG--XY,
for a sample with linear size $L=160$.
The value $\mu=-1$ was used in this simulation.
}
\label{d2n3pair}
\end{figure}

\begin{figure}[h!]
\centerline{
\resizebox{\columnwidth}{!}{\includegraphics{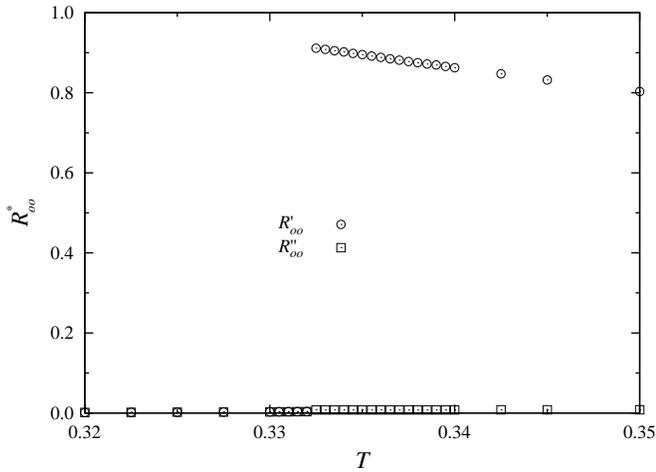}}}
\caption{
Simulation estimates for the quantities $R^*_{oo}$, 
obtained with $L=160$ for the two--dimensional LG--XY and $\mu=-1$.
}
\label{d2n3rstar}
\end{figure}

Simulation results
for the pair occupation probabilities, reported in Fig.
\ref{d2n3pair} and the excess quantities $R_{oo}^*$ shown in Fig.
\ref{d2n3rstar}, reveal that these quantities are discontinuous at the
first order transition region. On the other hand they show the effects
caused by the ferromagnetic interaction on the density in the system.
The quantity $R_{oo}''$ remains negligible due to the absence of
purely positional interaction. The behaviour of these quantities
follow in general the trends of the mean Hamiltonian and the density.
To summarize we found that  the system exhibits a first order phase
transition form a 
dense BKT phase to a paramagnetic one; 
in the temperature--density  phase diagram,
both phases are expected to coexist over some range of densities and
temperatures.

\subsection{$D=3$ and first--order transitions}
Simulation results
presented in  this subsection for the 
three--dimensional PR
and He models show the effects caused by large negative $\mu$ on
their transitional  behaviour, and, on the other hand, 
can be used to check the
predictions of the molecular--field like 
treatments used to  
construct the phase diagrams reported in our previous papers
\cite{romano2000,chamati2005a}; we refer to them
for further details, and present here  only the final
numerical results for the specific cases of interest.

It is  well known that these approximate treatments 
do not describe fluctuations
adequately, so that their predictions have to be taken with caution.
For example, MF predicted a
first order phase transition at $\mu=0$, while TSC and MC gave
evidence of a second order phase transition for He
\cite{chamati2005a}. For both three--dimensional models, simulations
performed for a selected value of the chemical potential,
revealed that MF describes qualitatively well the transitional properties
of the named models and that TSC improves upon it.
In Table
\ref{table03} we report results for the transition temperature
obtained, via simulations, for some values of $\mu$ for these models
so far. Here also one can read
that the transition temperature 
decreases with decreasing $\mu$.

\begin{table}
\caption[]{Transition temperatures $\Theta$ and ``critical''
particle density $\rho_c$ of PR and He models for some
selected values of the chemical potential $\mu$. Depending on $\mu$,
there is 
either a second order transition (II) or a first
order one (I); $\rho_c$ 
denotes the density at the
second order transition
temperature.}
\label{table03}
\begin{ruledtabular}
\begin{tabular}{lrcll}
Model &$\mu$\ \ & Transition  &   $\Theta$        &        $\rho_c$               \\
\hline
PR(n=2)&$\infty$&   II       &$2.201\pm0.003$                  &  1.            \\
      &$ 0.1$   &   II       &$1.423\pm0.003$     &$0.6900\pm0.004$             \\
      &$-1.5$   &   I        &$0.794\pm0.001$     &  $-$                        \\
He(n=3)&$\infty$&   II       &$1.443\pm0.001$\cite{peczak1991} &  1.            \\
      &$0.0$    &   II       &$0.998\pm0.001$\cite{chamati2005a}&
                                       $0.743\pm0.002$                          \\
      &$-1.5$   &   I        &$0.557\pm0.001$     &  $-$                        \\
\end{tabular}
\end{ruledtabular}
\end{table}

\begin{figure}[h!]
\centerline{
\resizebox{\columnwidth}{!}{\includegraphics{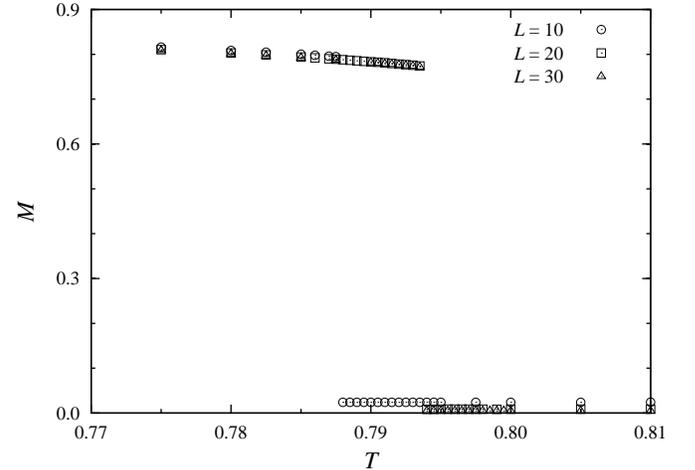}}}
\caption{
Simulation results for the magnetisation $M$ against the temperature for
the three--dimensional LG--PR,
obtained with different sample sizes and  $\mu=-1.5$.
}
\label{d3n2mag}
\end{figure}

Simulation results for both models exhibited a recognizable qualitative
similarity, so that only plots of PR are presented here. 
Behaviours of
observables such as  mean energy, density $\rho$ and  magnetisation $M$
(shown in Fig. \ref{d3n2mag}) were found to be either size independent
or to depend slightly on sample sizes in the transition region.
Furthermore, for all examined sample sizes,
we found abrupt jumps of these observables,
taking place over a narrow temperature range, $\Delta T =0.0005$. 

In Table \ref{table04} we present the transitional 
properties such as jumps in mean energy per site and density,
respectively, as well as magnetisation in the ordered phase, at the first order
phase transition undergone by the three--dimensional PR and He;
these results were obtained via MC,
MF and TSC. Comparison shows that TSC produces  a better estimate 
than MF for the transition temperature; on the other hand,
MF better predicts the jumps of thermodynamic quantities at the transition. 
In general,
according to the results gathered in Table \ref{table03}, one can see
that the phase diagram predicted by the approximate molecular field
theories is at least qualitatively correct. This fact is confirmed by
the recent simulation results for the phase diagram of the
diluted PR reported in Ref. \cite{maciolek2004}.

\begin{table*}
\caption[]{
Estimates for some properties at first--order
transition for the three--dimensional PR and He obtained by different
approaches. The results are obtained with $\mu=-1.5$.}
\label{table04}
\begin{ruledtabular}
\begin{tabular}{llllll}
Model & Method &  $\Theta$ & $\Delta H$ & $\Delta \rho$ & $M$ 
\\
\hline
PR & MC &$0.794\pm0.001$&$0.910\pm0.004$&$0.684\pm0.002$&$0.772\pm0.002$  \\
   & MF &   $0.741$     &    $1.138$    &  $0.849$      & $0.897$         \\
   &TSC &   $0.760$     &    $1.518$    &  $0.756$      & $0.903$         \\
He & MC &$0.557\pm0.001$&$0.882\pm0.003$&$0.877\pm0.001$&$0.804\pm0.001$  \\
   & MF &   $0.462$     &    $0.944$    &  $0.958$      & $0.9126$        \\
   &TSC &   $0.482$     &    $0.959$    &  $0.786$      & $0.888$         \\
\end{tabular}
\end{ruledtabular}
\end{table*}

\begin{figure}[h!]
\centerline{
\resizebox{\columnwidth}{!}{\includegraphics{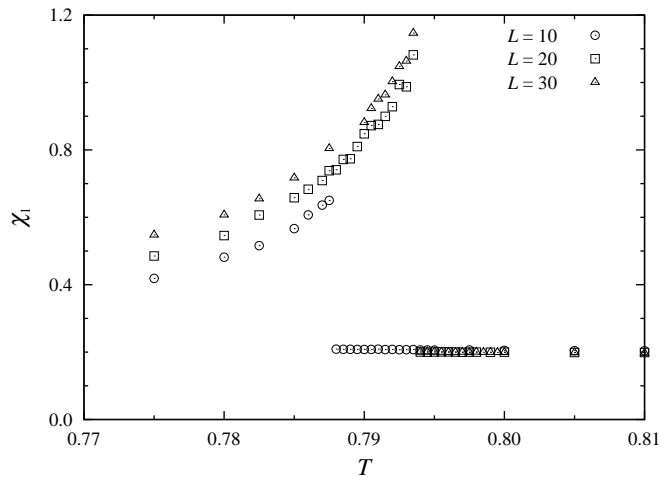}}}
\caption{
Simulation results for the susceptibility $\chi_1$ for the 
three--dimensional LG--PR, obtained with
different sample sizes. The associated statistical errors, not shown,
range up to $10\%$.
The value $\mu=-1.5$ was used in this simulation.
}
\label{d3n2chi}
\end{figure}

The susceptibility, actually
$\chi_1$, reported in Fig. \ref{d3n2chi}, showed a peak at a
temperature about 0.792, a strong 
sample size dependence below this
temperature and no sensitivity  to the sample sizes above it. The
behaviours of the 
three derivatives $C_{\mu V},~\rho_T,~\rho_{\mu}$ (not reported)
were found to be 
{\em qualitatively} similar to Fig. \ref{d2n2mu150der}).

\begin{figure}[h!]
\centerline{
\resizebox{\columnwidth}{!}{\includegraphics{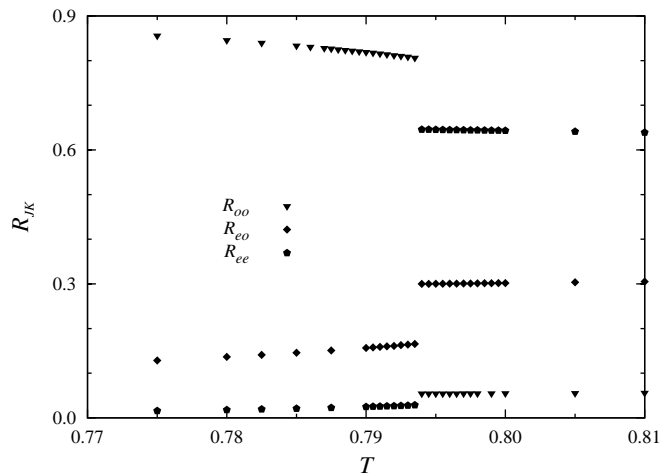}}}
\caption{
Simulation results for the three pair occupation probabilities
$R_{JK}$ obtained for the  three--dimensional LG--PR with linear sample size
$L=30$. The value $\mu=-1.5$ was used in this simulation.
}
\label{d3n2pair}
\end{figure}

Other fluidlike quantities such as  $R_{JK}$ (Fig \ref{d3n2pair}) and
$R_{oo}^*$ (not reported here) show how the density behaves when
the three--dimensional PR lattice gas model exhibits a first order
transition. These quantities are discontinuous at the transition
temperature and follow the behaviour obtained for the density and the
mean energy; once more we witnessed the smallness of the excess quantities
$R_{oo}$ due to the absence of purely positional interaction.
{In general we remarked a pronounced qualitative
similarity between the behaviours of the fluidlike quantities in
the present case and those discussed in the previous subsection for 2D
models.}

\section{Concluding remarks}\label{conclusion}

We have studied the critical properties of four 
LG  models
defined by $\mu<0$ and sufficiently large in magnitude,
$\mu=-D/2$, plus an additional case ($D=2$, PR, $\mu=-3D/4$).
This allowed us to investigate the impact of the chemical
potential on the nature of phase transition of these models and thus
to gain insights into their phase diagrams. Our simulations were
performed in the absence of pure positional interaction. A number of
thermodynamic quantities including some characteristics of fluid
systems were estimated. 
It was found that the common feature of most cases 
is the onset of a first--order phase transition
induced by the ferromagnetic interaction, 
and where an an abrupt change in the density of the
system was observed.

In two--dimensions we have investigated 
{both 
PR and XY models for $\mu=-D/2$.}
At this value of the chemical potential they showed
different critical behaviours. PR exhibited a BKT phase transition,
while XY showed a first order one. This might be a consequence of the
fact that the two models entail
different anchorings with respect to the horizontal plane in spin
space. 
{PR was further studied for  $\mu=-3D/4$, where 
evidence of a first order transition was found.}
{The change of the nature of the phase transition from
BKT to a discontinuous one}
{agrees with} 
previous RG
predictions \cite{rhe01,rhe03} and rigorous mathematical results
\cite{rERZ}; on the other hand, in recent simulation studies 
of quenched dilution it was found that the
transition temperature vanishes  
below the percolation threshold 
\cite{leonel2003,berche2003,surungan2005,wysin2005}.

Notice that usage of the Grand--Canonical
ensemble allows quite wide changes of density with temperature.
Such changes are obviously excluded from the start in the treatment of a 
quenched--dilution model.
Thus, there are significant differences between both methods,
yet the two 
resulting pictures are somehow compatible.

{Phase transition and critical dynamics
in site--diluted arrays of Josephson junctions were recently
studied experimentally in Ref. \cite{yun2006}; according
to the Authors' results, 
the BKT transition is altered by the introduction of percolative 
disorder far below the percolation threshold. 
Furthermore, the Authors of Ref. \cite{yun2006} 
found evidences of  a non--BKT--type superconducting transition for 
strongly disordered samples, taking place at finite temperature.
 Our results suggest that the 
transition in the named region becomes  of first order.}

For the three dimensional models
investigated here, i.e. PR and He, 
we found a first order phase transition form a ferromagnetic dense
phase to a diluted paramagnetic one.
The results obtained via simulation for $\mu=-D/2$ were found to
confirm those obtained by molecular field approximations used to
construct the phase diagram of Refs. \cite{romano2000,chamati2005a},
showing that the phase diagrams obtained there are qualitatively correct.

\acknowledgments
The present calculations were carried out, on, among other machines,
workstations belonging to the Sezione di Pavia of INFN (Istituto
Nazionale di Fisica Nucleare). Allocation of computer time by the
Computer Centre of Pavia University and CILEA (Consorzio
Interuniversitario Lombardo per l' Elaborazione Automatica, Segrate -
Milan), as well as by CINECA (Centro Interuniversitario Nord-Est di
Calcolo Automatico, Casalecchio di Reno - Bologna), are gratefully
acknowledged as well.
H. Chamati also acknowledges financial support from
Grant No. BK6/2007 of ISSP-BAS.
The authors also thank Prof. V. A. Zagrebnov (CPT--CNRS and
Universit\'e de la M\'editerran\'ee, Luminy, Marseille, France) and
Prof. A. C. D. van Enter (Rijksuniversiteit Groningen , the Netherlands)
for helpful discussions.

\end{document}